\RequirePackage{fix-cm}
\documentclass[twocolumn,epjc3]{svjour3}  
\smartqed  
\RequirePackage{graphicx}
\usepackage{multirow}
\usepackage{xcolor}
\usepackage{amsmath}
\usepackage{amssymb}
\usepackage{hyperref}
\usepackage{microtype} 
\usepackage{booktabs}
\usepackage{multirow}
\usepackage{subcaption}
\usepackage{url}

\usepackage[numbers,square,comma,sort&compress]{natbib}

\usepackage{lineno}
\journalname{Eur. Phys. J. C}
\begin{document}
%
\title{
Improved background modeling for dark matter search with COSINE-100
}



\author{
  G.H. Yu\thanksref{e1,addr1,addr2} \and
  N. Carlin\thanksref{addr4} \and
  J.Y. Cho\thanksref{addr2,addr12} \and
  J.J. Choi\thanksref{addr5,addr2} \and
  S. Choi\thanksref{addr5} \and
  A.C. Ezeribe\thanksref{addr6} \and
  L.E. Fran\c{c}a\thanksref{addr4} \and
  C. Ha\thanksref{addr7} \and
  I.S. Hahn\thanksref{addr8,addr9,addr10} \and
  S.J. Hollick\thanksref{addr3} \and
  E.J. Jeon\thanksref{e2,addr2,addr10} \and
  H.W. Joo\thanksref{addr5} \and
  W.G. Kang\thanksref{addr2} \and
  M. Kauer\thanksref{addr11} \and
  B.H. Kim\thanksref{addr2} \and
  H.J. Kim\thanksref{addr12} \and
  J. Kim\thanksref{addr7} \and
  K.W. Kim\thanksref{addr2} \and
  S.H. Kim\thanksref{addr2} \and
  S.K. Kim\thanksref{addr5} \and
  W.K. Kim\thanksref{addr10,addr2} \and
  Y.D. Kim\thanksref{addr2,addr13,addr10} \and
  Y.H. Kim\thanksref{addr2,addr14,addr10} \and
  Y.J. Ko\thanksref{addr2} \and
  D.H. Lee\thanksref{addr12} \and
  E.K. Lee\thanksref{addr2} \and
  H. Lee\thanksref{addr10,addr2} \and
  H.S. Lee\thanksref{addr2,addr10} \and
  H.Y. Lee\thanksref{addr8} \and
  I.S. Lee\thanksref{addr2} \and
  J. Lee\thanksref{addr2} \and
  J.Y. Lee\thanksref{addr12} \and
  M.H. Lee\thanksref{addr2,addr10} \and
  S.H. Lee\thanksref{addr10,addr2} \and
  S.M. Lee\thanksref{addr5} \and
  Y.J. Lee\thanksref{addr7} \and
  D.S. Leonard\thanksref{addr2} \and
  N.T. Luan\thanksref{addr12} \and
  B.B. Manzato\thanksref{addr4} \and
  R.H. Maruyama\thanksref{addr3} \and
  R.J. Neal\thanksref{addr6} \and
  S.L. Olsen\thanksref{addr2} \and
  B.J. Park\thanksref{addr10,addr2} \and
  H.K. Park\thanksref{addr15} \and
  H.S. Park\thanksref{addr14} \and
  J.C. Park\thanksref{addr16} \and
  K.S. Park\thanksref{addr2} \and
  S.D. Park\thanksref{addr12} \and
  R.L.C. Pitta\thanksref{addr4} \and
  H. Prihtiadi\thanksref{addr17} \and
  S.J. Ra\thanksref{addr2} \and
  C. Rott\thanksref{addr1,addr18} \and
  K.A. Shin\thanksref{addr2} \and
  D.F.F.S. Cavalcante\thanksref{addr4} \and
  M.K. Son\thanksref{addr16} \and
  N.J.C. Spooner\thanksref{addr6} \and
  L.T. Truc\thanksref{addr12} \and
  L. Yang\thanksref{addr19} \\
  (COSINE-100 Collaboration)
}

\thankstext{e1}{e-mail: tksxk752@naver.com}
\thankstext{e2}{e-mail: ejjeon@ibs.re.kr (Corresponding author)}

\institute{Department of Physics, Sungkyunkwan University, Suwon 16419, Republic of Korea \label{addr1}
  \and
  Center for Underground Physics, Institute for Basic Science (IBS), Daejeon 34126, Republic of Korea \label{addr2}
  \and
  Physics Institute, University of S\~{a}o Paulo, 05508-090, S\~{a}o Paulo, Brazil \label{addr4}
  \and
  Department of Physics, Kyungpook National University, Daegu 41566, Republic of Korea \label{addr12}
  \and
  Department of Physics and Astronomy, Seoul National University, Seoul 08826, Republic of Korea \label{addr5}
  \and
  Department of Physics and Astronomy, University of Sheffield, Sheffield S3 7RH, United Kingdom \label{addr6}
  \and
  Department of Physics, Chung-Ang University, Seoul 06973, Republic of Korea \label{addr7}
  \and
  Center for Exotic Nuclear Studies, Institute for Basic Science (IBS), Daejeon 34126, Republic of Korea \label{addr8}
  \and
  Department of Science Education, Ewha Womans University, Seoul 03760, Republic of Korea \label{addr9}
  \and
  IBS School, University of Science and Technology (UST), Daejeon 34113, Republic of Korea \label{addr10}
  \and
  Department of Physics and Wright Laboratory, Yale University, New Haven, CT 06520, USA \label{addr3}
  \and
  Department of Physics and Wisconsin IceCube Particle Astrophysics Center, University of Wisconsin-Madison, Madison, WI 53706, USA \label{addr11}
  \and
  Korea Research Institute of Standards and Science, Daejeon 34113, Republic of Korea \label{addr14}
  \and
  Department of Accelerator Science, Korea University, Sejong 30019, Republic of Korea \label{addr15}
  \and
  Department of Physics and IQS, Chungnam National University, Daejeon 34134, Republic of Korea \label{addr16}
  \and
  Department of Physics, Universitas Negeri Malang, Malang 65145, Indonesia \label{addr17}
  \and
  Department of Physics and Astronomy, University of Utah, Salt Lake City, UT 84112, USA \label{addr18}
  \and
  Department of Physics, University of California San Diego, La Jolla, CA 92093, USA \label{addr19}
}

\date{Received: date / Accepted: date}

\date{Received: date / Accepted: date}

\maketitle

\begin{abstract}
\sloppy
COSINE-100 aims to conclusively test the claimed dark matter annual modulation signal detected by DAMA/LIBRA collaboration. DAMA/LIBRA has released updated analysis results by lowering the energy threshold to 0.75\,keV through various upgrades. They have consistently claimed to have observed the annual modulation.
In COSINE-100, it is crucial to lower the energy threshold for a direct comparison with DAMA/LIBRA, which also enhances the sensitivity of the search for low-mass dark matter, enabling COSINE-100 to explore this area.
Therefore, it is essential to have a precise and quantitative understanding of the background spectrum across all energy ranges. 
This study expands the background modeling from 0.7 to 4000\,keV using 2.82\,years of COSINE-100 data. 
The modeling has been improved to describe the background spectrum across all energy ranges accurately.
Assessments of the background spectrum are presented, considering the nonproportionality of NaI(Tl) crystals at both low and high energies and the characteristic X-rays produced by the interaction of external backgrounds with materials such as copper. Additionally, constraints on the fit parameters obtained from the alpha spectrum modeling fit are integrated into this model. These improvements are detailed in the paper.
\end{abstract}  

\section{Introduction}
\label{sec:intro}
There have been many attempts to detect dark matter particles known as Weakly Interacting Massive Particles (WIMPs) by observing the recoil of nuclei resulting from interactions between WIMPs and nuclei~\cite{Gaitskell:2004gd,particle2020review}. 
However, except for the DAMA/LIBRA experiment, no other evidence has been found so far. The DAMA/LIBRA experiment uses an array of NaI(Tl) crystals to observe the annual modulation phenomenon of dark matter. They have consistently observed annual modulation signatures~\cite{DAMA:2008jlt,DAMA:2010gpn,Bernabei:2013xsa}. According to their findings, the model-independent annual modulation signature results were obtained with a 9.6~$\sigma$ C.L. in the energy range of 1 to 6\,keV in Phase~2~\cite{Bernabei:2018jrt}. Furthermore, they have lowered the energy threshold to 0.75\,keV and are still reporting consistent results. However, DAMA's signal conflicts with the null results of other experiments assuming the standard halo model of dark matter~\cite{bernabei2023dark}.

COSINE-100 is an experiment to detect annual modulation using an array of 106 kg NaI(Tl) crystals to test the DAMA/LIBRA signal. COSINE-100 has tested on the DAMA/LIBRA results using model-dependent~\cite{adhikari2021strong} and model-independent~\cite{adhikari2022three} methods with a 1\,keV threshold. However, to directly test the DAMA/LIBRA's new 0.75\,keV threshold, COSINE-100 needs to lower the energy threshold to below 1\,keV. 
By lowering the energy threshold, the sensitivity of the search for low-mass dark matter is also improved, allowing COSINE-100 to explore this area more effectively. 
To achieve this, a more accurate background model, which encompasses the low-energy region, is needed.
Some radioisotopes, such as $^{22}$Na, produce both low- and high-energy emissions associated with each other. 
Thus, it is important to develop a more accurate background model that encompasses both low- and high-energy regions, including the 3–4 MeV region. 
Moreover, a comprehensive understanding of the high-energy regions can be utilized for detecting boosted dark matter signals in COSINE-100~\cite{adhikari2023search}.

We have developed a background model by using 2.82 years of COSINE-100 data. The model was expanded to cover the energy range from 0.7 to 4000\,keV, providing a more concrete and accurate representation of the background spectrum. 
In developing this model, we performed the energy calibration that accounts for the nonproportionality of the NaI(Tl) crystals, as studied in~\cite{COSINE-100:nPR}. 
Since the MC simulations do not implement a nonproportional scintillation response, we incorporated the calibration curve derived from the experimental data into the simulated background spectrum. This ensures that the simulated spectra reflect the true energy response of the NaI(Tl) crystals, enhancing the accuracy of the background model. The detailed methodology is described in Sections ~\ref{sec:anode_calibration} and \ref{sec:dynode_calibration}.

The $\gamma$/X-rays emitted by background sources can excite atoms in copper or materials close to crystal detectors, producing X-rays with energies in the few keV range. These X-rays contribute to the backgrounds observed at low energies and are included in this study to improve modeling accuracy.
Additionally, we conducted a separate study to model the alpha spectrum~\cite{COSINE-100:alpha}. We integrated the constraints from the alpha model fit to refine the current approach. The details are provided in Section~\ref{sec:modeling}.

\section{The COSINE-100 experiment}
\label{sec:experiment}
The COSINE-100 experiment uses NaI(Tl) crystals like DAMA/LIBRA. Eight crystals weighing 8 to 18 kg each are aligned in two layers and immersed in a 2200\,L of linear alkylbenzene (LAB)-based liquid scintillator (LS) acting as an active/passive shield. The LS is contained in an acrylic box container, surrounded by a 3\,cm thick copper box, a 20\,cm thick lead-brick castle, and plastic scintillator panels to shield it from external backgrounds. 
The NaI(Tl) crystal used in the experiment is cylindrical and wrapped with about ten layers of 250\,$\mu$m thick Teflon. It is optically coupled with a 12.0\,mm thick quartz window at both ends, with a 1.5\,mm thick optical pad in between. This is encapsulated with 1.5\,mm thick OFE copper. The quartz at both ends of the crystal encapsulator is optically coupled with 3-inch Hamamatsu R12669SEL  photomultiplier tubes (PMTs) using a small amount of high-viscosity optical grease~\cite{Adhikari:2017esn}.
The experiment started on September 22, 2016, and ran until April 2023. We used the data collected for 2.8\,years until June 2019 to improve the background modeling.

To verify the recent results of DAMA/LIBRA, it is necessary to lower the energy threshold below 1\,keV and ensure that the background level in the WIMP region of interest (ROI) below 6\,keV is comparable to the 1 to 2\,counts/keV/day (dru) level. Our previous COSINE-100 background modeling has shown that the average background level in the (1--6)\,keV region is 2.85$\pm$0.15\,dru~\cite{COSINE-100:2021mrq}. To improve the understanding of backgrounds, we need to lower the energy threshold while still attaining high event selection efficiency at low energy.
The crystals' effective light yield was estimated to be 15 photoelectrons per keV using the 59.6\,keV $\gamma$ ray from $^{214}$Am, which is more than twice that of DAMA/LIBRA. 
However, below 2 keV, the scintillation events are contaminated by PMT noise events.~\cite{COSINE-100:2020wrv}. As the energy decreases, it becomes increasingly challenging to separate the noise events, as there are only a few pulses. To address this issue, we implemented a multi-layer perceptron (MLP) network based on noise-separating pulse-shape discriminating (PSD) parameters. As a result, we were able to achieve a threshold of 0.7~keV with 20\% efficiency for event selection and less than 1\% noise contamination~\cite{COSINE-100:eventselection-2024}.
This can enhance background modeling in the extended energy region down to 0.7\,keV.

In the low-energy region, there is the influence of not only $^{3}$H and $^{210}$Pb but also peaks of 0.87\,keV from $^{22}$Na and 3.2\,keV from $^{40}$K.
A better understanding of their contributions can be achieved by analyzing the background distribution extended to 0.7\,keV.
This analysis can help achieve more accurate modeling by considering the crystal's nonproportionality and applying it to the background obtained through Monte Carlo simulation based on Geant4. Therefore, the following section discusses energy calibration while considering nonproportionality.

\section{Energy calibration}
\label{calibration}
\subsection{Energy calibration and energy resolution in low energy region}
\label{sec:anode_calibration}
It is well-known that the relationship between the energy of an incident gamma ray and the amount of light produced in scintillation detectors such as NaI(Tl) is not linear~\cite{engelkemeir1956nonlinear,jones1962nonproportional,leutz1997scintillation,collinson1963fluorescent,khodyuk2010nonproportional,moses2012origins}. 
This nonproportionality in the ROI of dark matter search should be considered carefully, and it is essential to account for it when calibrating the gamma energy accurately. 
Previous calibration studies of COSINE-100 have reported about 20\% nonproportionality at the 1~keV level compared to the 50~keV energy response~\cite{COSINE-100:2021mrq}. As this could significantly impact WIMP detection, a thorough understanding of the detector response, including the extended threshold region, is necessary.
To achieve this, we used internal peaks from the NaI(Tl) crystals of COSINE-100 as calibration points. The internal peaks include 0.87\,keV from $^{22}$Na, 3.2\,keV from $^{40}$K, 25\,keV from $^{109}$Cd, 28\,keV from $^{113}$Sn, 39\,keV from $^{125}$I, 49\,keV from $^{210}$Pb, 67\,keV from $^{125}$I, and 88\,keV from $^{109}$Cd. 
To account for the nonproportionality, each peak was modeled with an integrated charge from the PMT charge output signal (waveform), and the relative light yield was measured as a function of the incident gamma energy. A detailed study on this is described in reference~\cite{COSINE-100:nPR}. The calibration curves consider the nonproportionality of five NaI(Tl) crystals (C2, C3, C4, C6, and C7), excluding three crystals due to a high noise rate and low light yields. These curves are applied to the simulated background spectrum of each crystal.

\begin{figure*}[h]
    \centering
    \centerline{\includegraphics[width=0.95\linewidth]{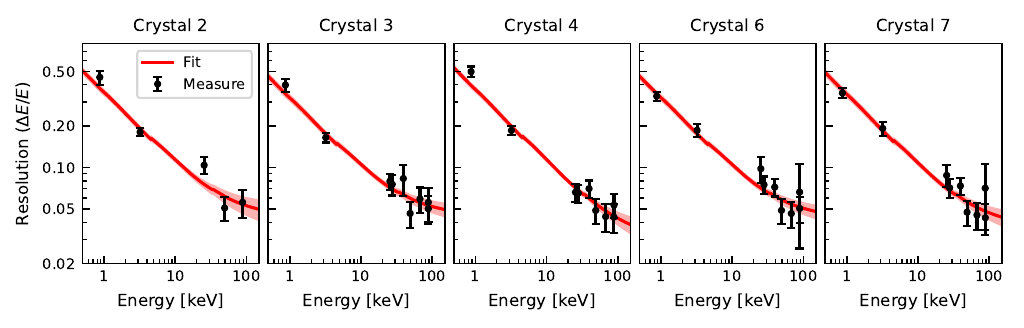}}
    \caption{
    Energy-dependent resolution used for each crystal. The resolution fit is based on internal peaks from the NaI(Tl) crystal. 
    }
    \label{fig:calibfunc}  
\end{figure*}

Another important parameter of the NaI(Tl) scintillation detector is energy resolution, crucial for dark matter search at low energies. It is usually expressed as the full width at half maximum of the peak and is affected by factors such as nonproportionality, crystals' inhomogeneity, and the Poisson fluctuation of the number of photoelectrons produced by the PMT. 
In the work published in~\cite{COSINE-100:nPR}, we modeled the energy resolution as a function of the incident gamma energy considering the above-mentioned effects. We validated it by conducting a waveform simulation~\cite{choi2024waveform}, which described the characteristics of the data well and in good agreement, including the direct measurement of light yield at 0.87\,keV. This allows us to apply the resolution model to the lower energy region. 
Figure~\ref{fig:calibfunc} presents the resolution curve used to smear the simulated background spectrum.

\subsection{PMT charge asymmetry correction}
\label{sec:asymcorr}
\begin{figure*}[h]
    \centering
  \begin{subfigure}{0.49\textwidth}
    \includegraphics[width=\linewidth]{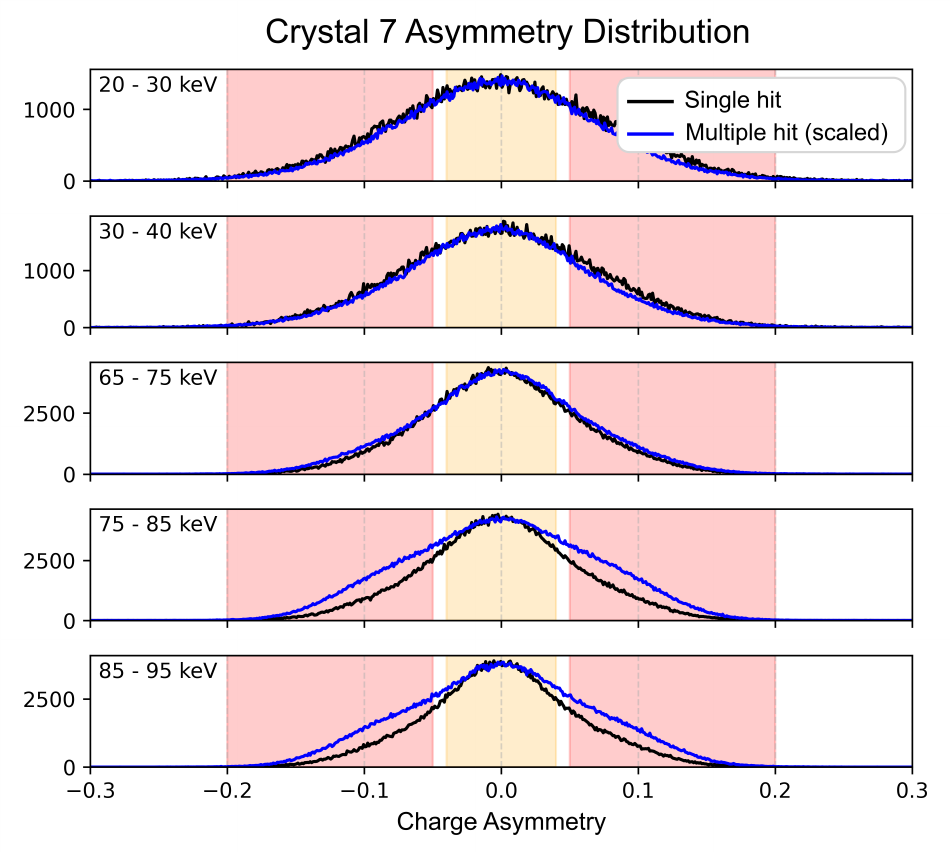}
    \caption{} 
  \end{subfigure}%
  \begin{subfigure}{0.49\textwidth}
    \includegraphics[width=\linewidth]{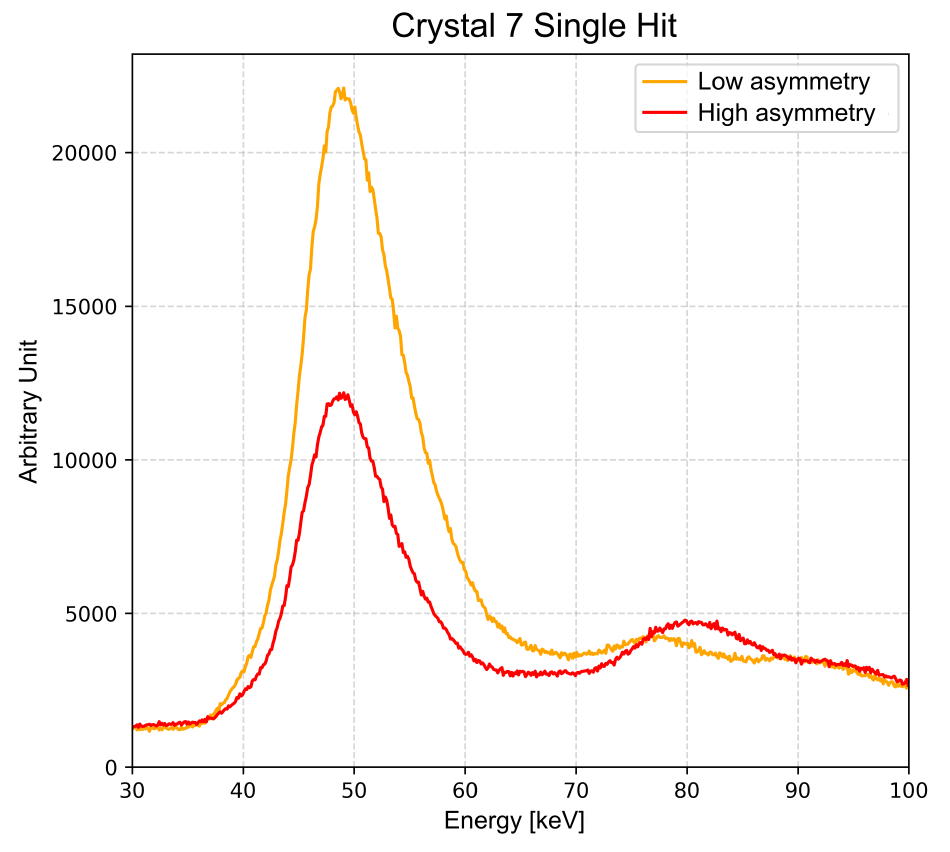}
    \caption{} 
  \end{subfigure}%
    \caption{(a) Comparison of the PMT charge asymmetry between single- and multiple-hit events in various energy intervals. The difference between the distributions is prominent only in the single-hit events above 75\,keV. (b) The energy spectrum of single-hit events for large and small charge asymmetry. Each small asymmetry and large asymmetry corresponds to the yellow and pink colored region in (a). 
    }
    \label{fig:pmtasym}  
\end{figure*}
When we compared the MC spectrum smeared with the energy-dependent resolution to the data, we found a discrepancy in the energy range of around 80\,keV for the single-hit events defined to have signals in only one of the crystals. 
According to the background simulations, events in this range are primarily due to the backgrounds from PMTs attached to the crystals. Specifically, there are X-ray deposits at 74.8 and 77.1\,keV from the decay of $^{214}$Pb in the $^{238}$U chain.
On the other hand, external backgrounds far from the crystals contribute to this energy range for multiple-hit events, and the MC spectrum agrees well with the data in this range.

This discrepancy is due to the asymmetry of the location of the background sources. If a background source from PMT contamination deposits its energy, more photons will be collected at the contaminated PMT than at the PMT on the opposite side. This discrepancy should also be incorporated into the Monte Carlo simulation. 
To verify this, we examined the PMT charge asymmetry between single-hit events and multiple-hit events at different energy intervals. The charge asymmetry is defined as,
\begin{equation}
Charge~Asymmetry = \frac{Q_1-Q_2}{Q_1+Q_2},  \\
\end{equation}
where $Q_1$ and $Q_2$ are the charges measured by each PMT attached to the crystal.
As shown in Fig.~\ref{fig:pmtasym} (a), while there is a noticable difference in charge asymmetry in the 75--95\,keV range between the single and multiple hit events, the difference is much smaller in the low energy region. We also compared the energy distribution of the events for the large charge asymmetry (pink area) and the small charge asymmetry (yellow area), as seen in Fig.~\ref{fig:pmtasym} (b). A noticeable change in spectral shape was observed for energies over 75\,keV, primarily due to the PMT contamination.
Based on these results, we corrected the MC spectrum for this charge asymmetry, which allows us to extend the low-energy fitting range from 70 to 90\,keV in the background modeling.

\subsection{Energy calibration in the high-energy region}
\label{sec:dynode_calibration}
We also evaluated nonproportionality at high energies~\cite{COSINE-100:nPR}. To do this, we measured the light yields of the prominent gamma peaks from the high-energy spectrum, which included 295\,keV from $^{214}$Pb, 511 and 1274\,keV from $^{22}$Na, 609 and 1764\,keV from $^{214}$Bi, 1461\,keV from $^{40}$K, and 2615\,keV from $^{208}$Tl. These measurements are in good agreement with previous measurement~\cite{DEVARE1963253}. 

In the background spectrum, the influence of $^{22}$Na and $^{208}$Tl is dominant above approximately 2000\,keV, and the nonproportionality of multiple gamma rays from the decay of $^{22}$Na or $^{208}$Tl contributes to the distortions in the spectral shape of the backgrounds. This is discussed in detail in the following section.

\subsubsection{$^{22}$Na energy calibration}
\label{sec:nacorr}
\begin{figure*}[ht]
    \centering
  \begin{subfigure}{0.325\textwidth}
    \includegraphics[width=\linewidth]{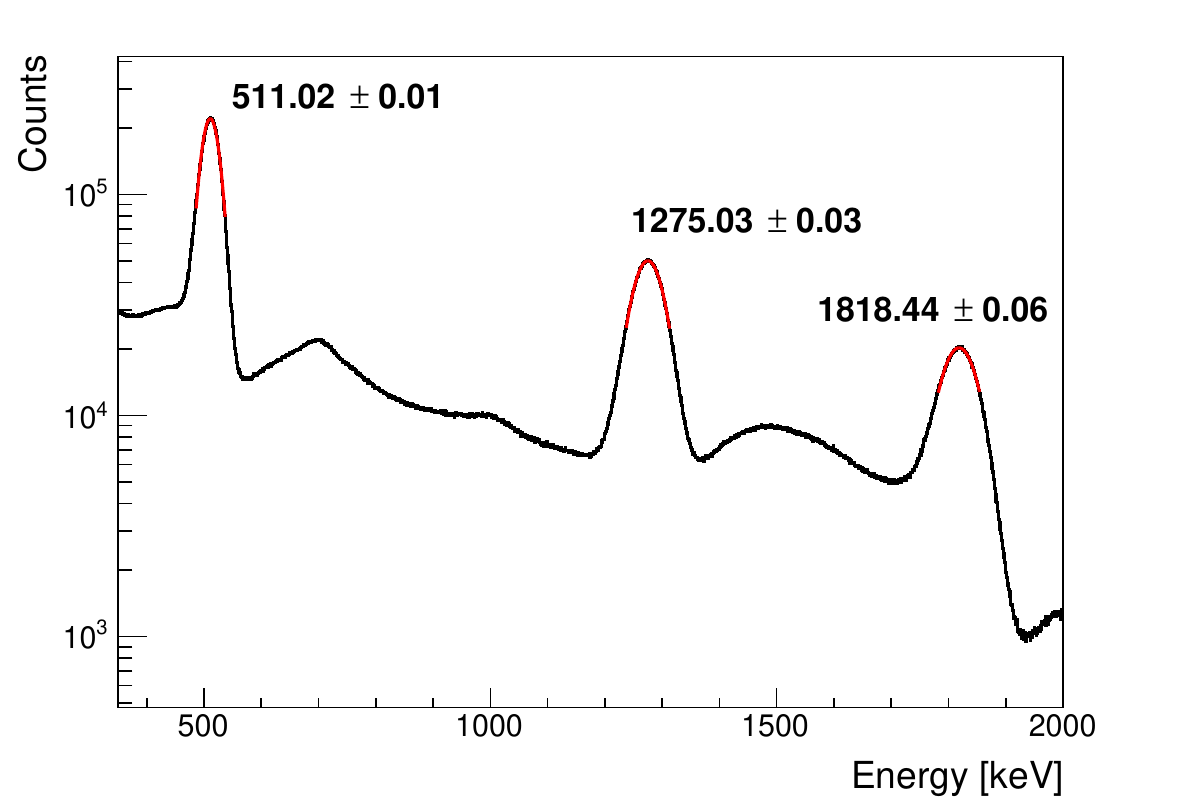}
    \caption{} 
  \end{subfigure}%
  \begin{subfigure}{0.325\textwidth}
    \includegraphics[width=\linewidth]{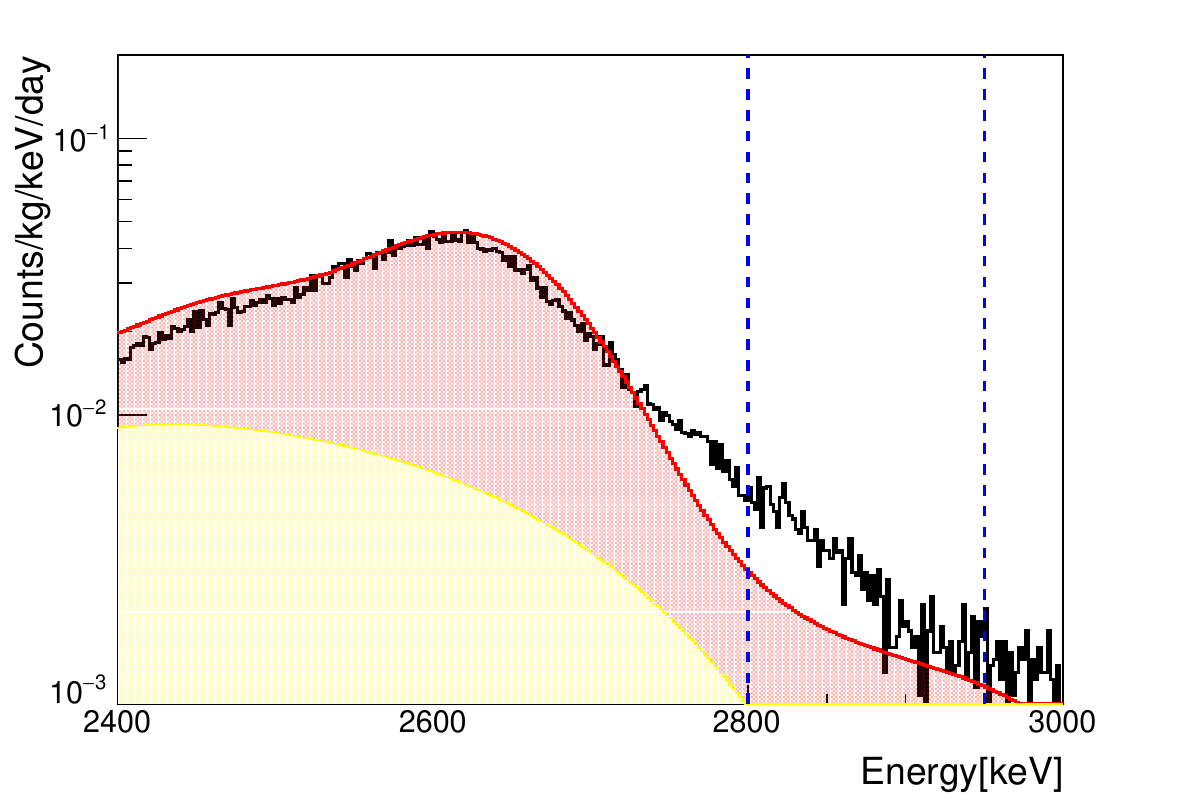}
    \caption{} 
  \end{subfigure}%
  \begin{subfigure}{0.325\textwidth}
    \includegraphics[width=\linewidth]{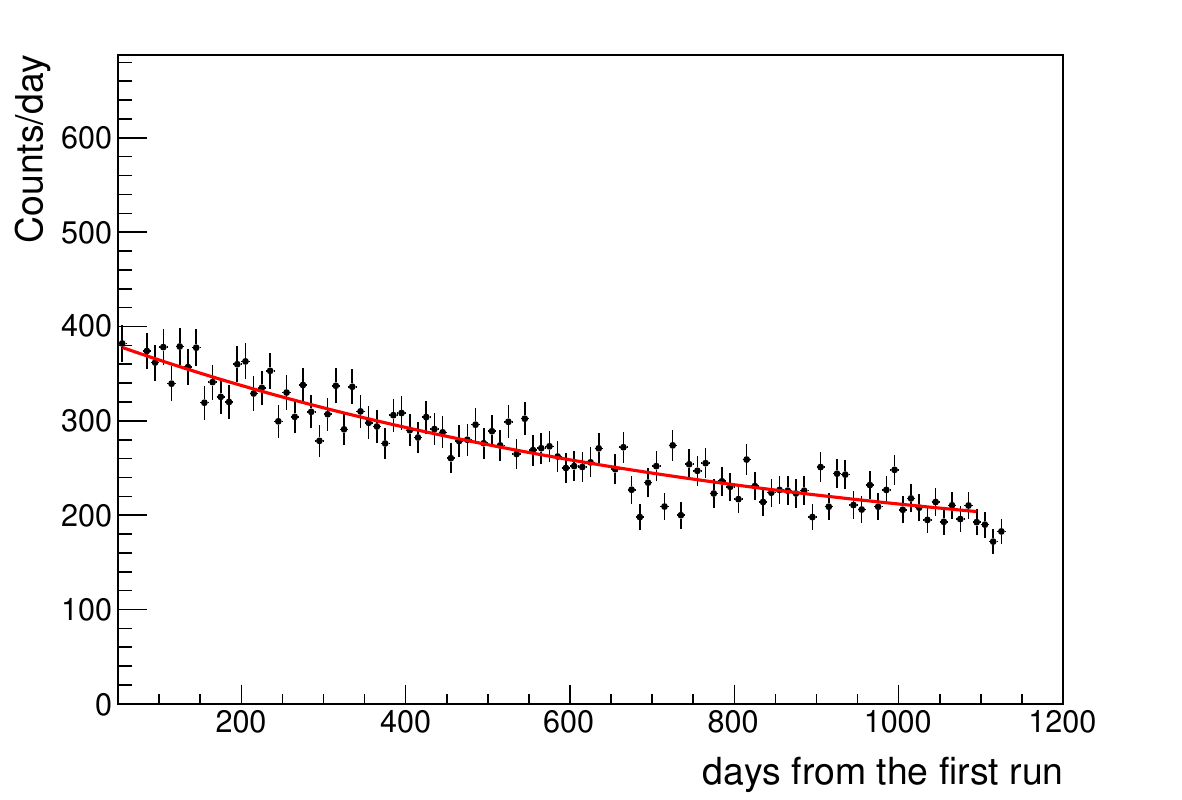}
    \caption{} 
  \end{subfigure}
    \caption{(a) Fitted energy peak values from $^{22}$Na source data. (b) Modeling result of C6 single-hit energy spectrum over a 1.7-year data period. Black lines represent the data, while $^{22}$Na and other MC modeling results are denoted by yellow and red regions, respectively. (c) Time spectrum showing count rates from all crystals in the 2800\,\mbox{--}\,2950\,keV energy range over a 3-year data period. 
    The dashed vertical lines in (b) indicate the corresponding energy range. The time-dependent data rate is fitted using an exponential function, resulting in a half-life of 3.04 $\pm$ 0.60 years.}
    \label{fig:dynode_22Na}  
\end{figure*}
\begin{figure*}[ht]
    \centering
  \begin{subfigure}{0.325\textwidth}
    \includegraphics[width=\linewidth]{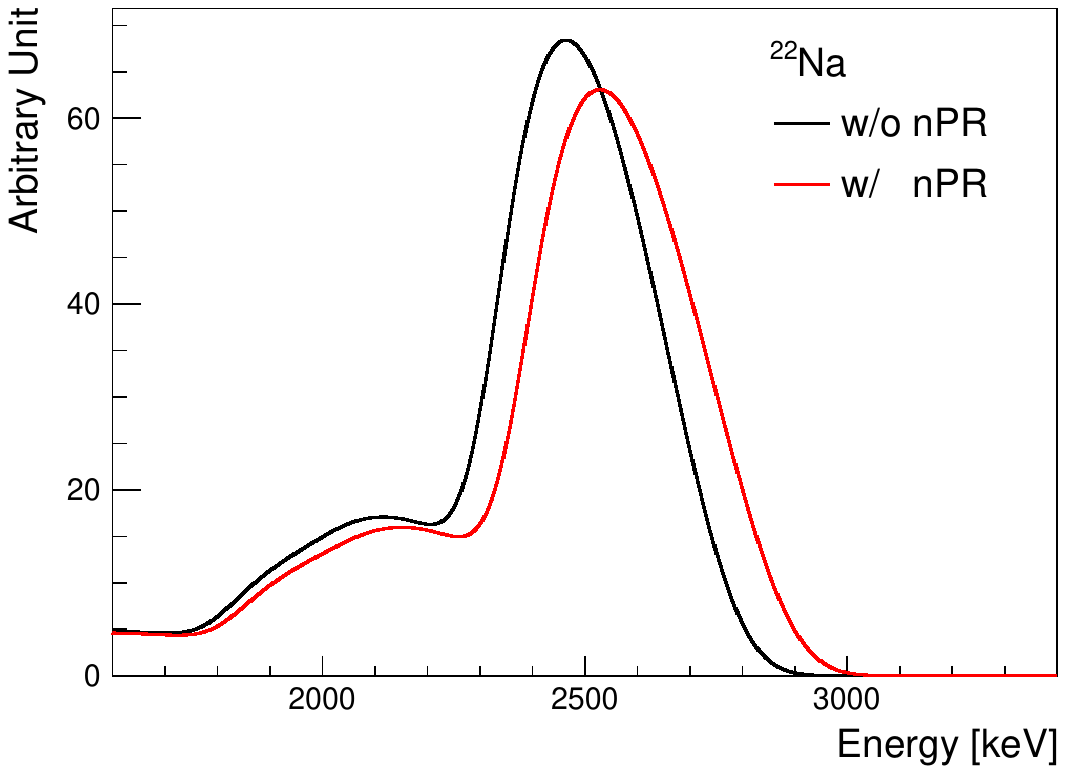}
    \caption{} \label{fig:1a}
  \end{subfigure}%
  \begin{subfigure}{0.325\textwidth}
    \includegraphics[width=\linewidth]{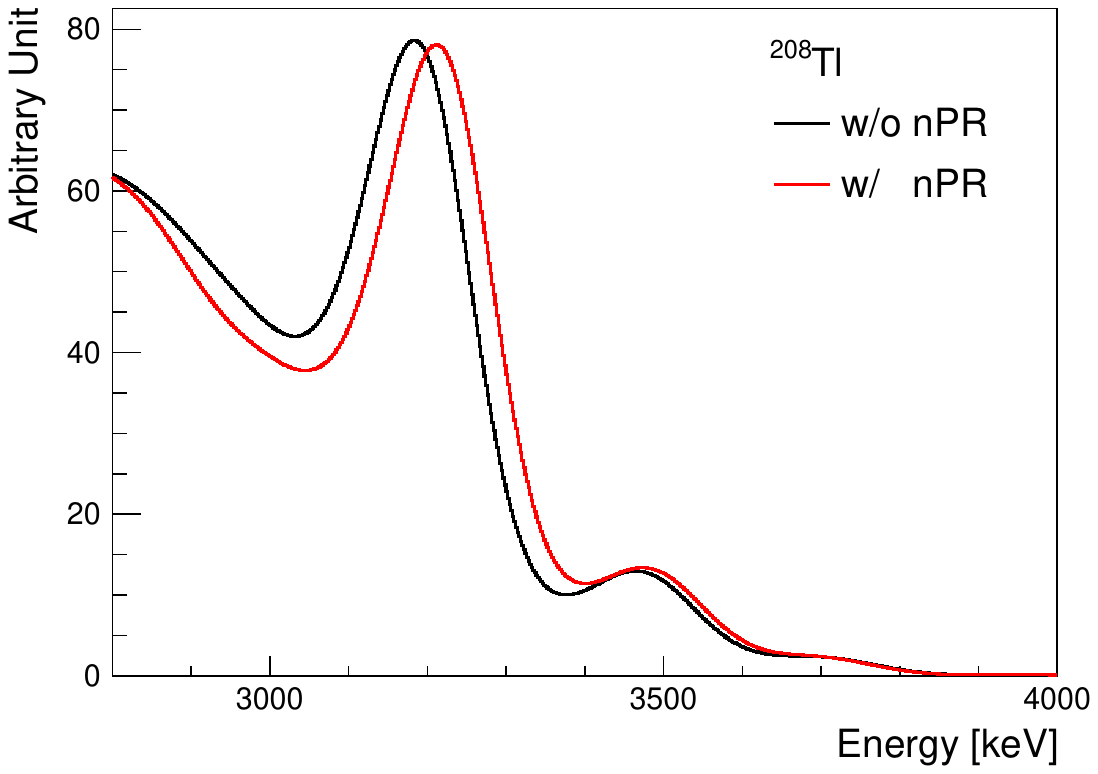}
    \caption{} \label{fig:1b}
  \end{subfigure}%
  \begin{subfigure}{0.325\textwidth}
    \includegraphics[width=\linewidth]{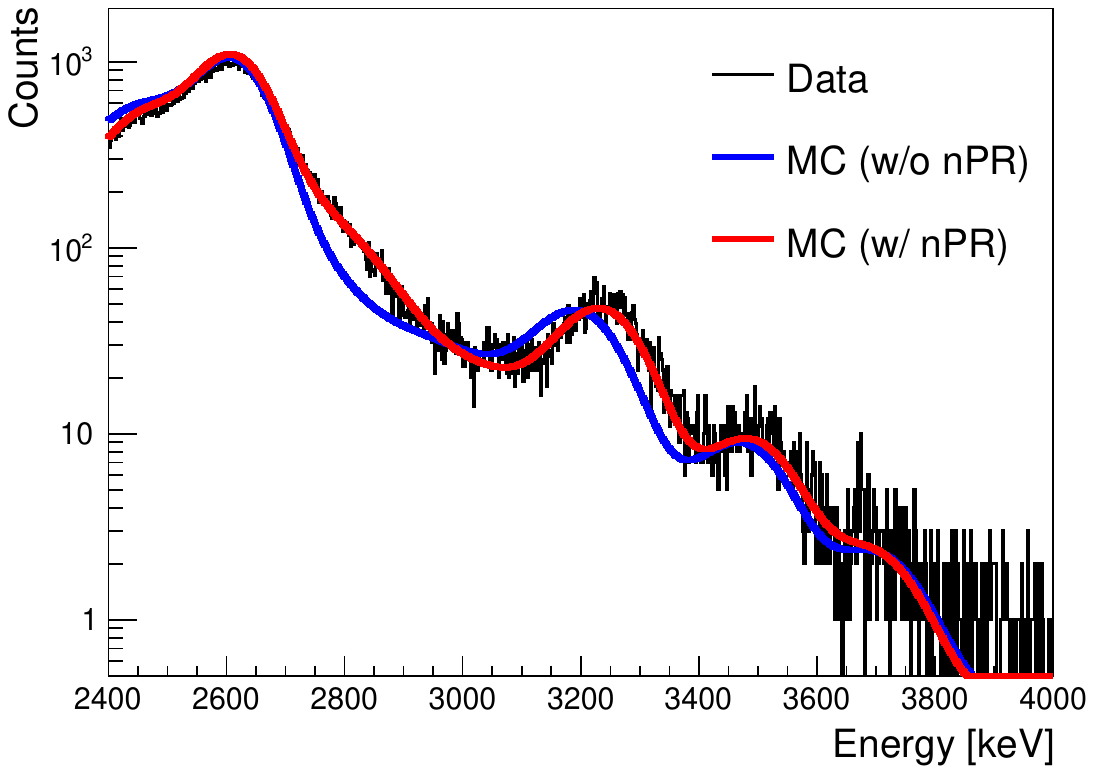}
    \caption{} \label{fig:1c}
  \end{subfigure}
    \caption{Comparison of $^{22}$Na (a) and $^{208}$Tl (b) with and without nonproportionality correction. (c) The modeling result with and without nonproportionality correction.}
    \label{fig:nPR_Dynode2}  
\end{figure*}

Approximately 90\% of $^{22}$Na decays to an excited state of $^{22}$Ne through the positron emission. This is followed by $^{22}$Ne$^{*}$ transitioning to the stable $^{22}$Ne isotope by releasing a gamma ray of 1274.6\,keV with a 5.3 ps mean lifetime.
The final-state positron immediately annihilates two 511\,keV gamma rays, and as a result, we expect three peaks at 511, 1274.6, and their sum of 1785.6\,keV. 
We examined these radiations using a $^{22}$Na source, by placing it near the COSINE-100 crystals and measured three peaks, as shown in Fig.~\ref{fig:dynode_22Na} (a).
It was observed that the sum peak position is not in agreement with the expected value and is located at around 1818\,keV. While the peaks of 511\,keV and 1275\,keV agree well with expectation. 
This discrepancy occurred because the combined nonproportionality effect on each 511 and 1274\,keV peak is greater than the nonproportionality effect on the 1786\,keV peak alone. This effect occurs in high-energy regions of the background spectrum, dominated by $^{22}$Na.
 
Figure~\ref{fig:dynode_22Na} (b) shows that the previous background modeling results do not fit the data well. When the event rate in the 2700-2950 keV region was evaluated over time and fitted with an exponential function, a fitted half-life of 3.04 $\pm$ 0.60 was obtained (Fig.~\ref{fig:dynode_22Na} (c)). This value, within its uncertainty, is consistent with that of $^{22}$Na (2.6029 $\pm$ 0.0008 years). The fit result indicates that the events in this region are from $^{22}$Na decay, but the energy spectrum is not well reproduced by the Monte Carlo simulation.
It is because multiple gamma rays are emitted during the $^{22}$Na decay, which is summed with the positron spectrum having an endpoint of 545\,keV and distributed up to a Q-value of 2842\,keV. In such cases, the summed background spectrum could be distorted by the combined nonproportionality effect. 

In Figure~\ref{fig:nPR_Dynode2} (a), we compared the contribution of $^{22}$Na with and without the effect of nonproportionality. We noticed an approximately 80\,keV shift in the sum peak for two 511\,keV gammas, 1274\,keV gamma, and positron emissions.
Similarly, $^{208}$Tl decay generates multiple gamma rays, such as 2614, 583, and 277\,keV, and the pileup energy is affected by approximately 20\,keV shift due to nonproportionality, as shown in Fig.~\ref{fig:nPR_Dynode2} (b).
As a result, Fig.~\ref{fig:nPR_Dynode2} (c) shows that the improved background modeling precisely describes the data 
in the high-energy region, expanding to 4000\,keV by considering nonproportionality.
This makes it possible to accurately determine the fractional contribution of the $^{22}$Na background and evaluate its impact at low energy. This is because 10\% of the $^{22}$Na decay occurs through EC capture, which results in X-rays of 0.87\,keV emitted from k-shell electron capture.

\section{Background modeling}
\label{sec:modeling}
We simulated contributions from all potential sources to assess the background spectrum in the modeling. These include internal backgrounds from the NaI(Tl) crystal detectors and external backgrounds from PMTs, greases, PTFE reflective sheets, copper encapsulators, bolts, cables, acrylic supports, liquid scintillator (LS), copper box, and a steel structure that supports the lead block housing. We evaluated background radiation caused by radioisotopes in the decay chains of $^{238}$U, $^{232}$Th, $^{40}$K, and$^{235}$U within the crystals and materials of the detector system included in the experimental enclosure. Furthermore, we evaluated the effects of surface contaminants and backgrounds from cosmogenically activated isotopes within the NaI(Tl) crystals.

To generate these background spectra, we utilized a COSINE-100-specific simulation framework based on Geant4~\cite{GEANT4:2002zbu,Allison:2006ve,Allison:2016lfl}, developed for modeling 59.5 days and 1.7 years of COSINE-100 data~\cite{COSINE-100:2018tfl,COSINE-100:2021mrq}. The simulated energy distributions are reconstructed by smearing them in a Gaussian shape based on the resolution function obtained from Sect.~\ref{calibration} and normalized by the measured activity. 
The data is fitted with the summed simulated spectrum to construct the background model, using the binned-likelihood fit method~\cite{COSINE-100:2021mrq}. During the fit, the amount of each background source is constrained within the uncertainties of the measured activity, allowing for floating unknown fractional activities. Detailed descriptions are provided in the following sections.

\subsection{Constraints on internal backgrounds}
\label{sec:internal}
For internal contaminations, radioactive $\alpha$-decay can be recognized by high-energy peaks in the background spectrum, corresponding to individual alpha peaks from the decays of $^{238}$U and $^{232}$Th. In a separate study~\cite{COSINE-100:alpha}, we analyzed alpha-alpha time-correlated events, by finding an $\alpha$ event followed by a delayed coincidence $\alpha$ event in a given time window. This analysis focused on decay sub-chains such as $^{228}$Th--$^{208}$Pb for internal $^{232}$Th, which are dominant at high energies. This method enables us to measure the activity levels of these sub-chains. In the modeling fit, these measurements constrain the fractional activity of $^{232}$Th backgrounds as prior information.
In addition, we used the measurement for $^{210}$Po to constrain the activity of $^{210}$Pb by assuming secular equilibrium of the decay chain. 
Another significant background component contaminating the crystal is the $^{40}$K isotope, which notably impacts the low-energy spectrum, resulting in X-ray emissions at 3\,keV. Its activity can be accurately measured using the 1460\,keV decay~\cite{Adhikari:2017esn}. This measurement is also used to constrain the fractional activity in the fit.

\subsection{Position dependence of background contaminants in PMTs} 
PMTs comprise three main components: the PMT window, body, and stem. Each component may have different levels of contamination due to the materials it is made of. However, measuring PMTs' activity is done collectively as assembled modules, reflecting a combined contribution from all three components. During the simulation of the background spectrum, variations in the position of background contaminants on these components resulted in differences in the spectral shape, particularly in the middle energy range of a few hundred keV.
Given PMTs' significant role as external background sources, their impacts are studied in the modeling process. PMT background simulations are conducted for each component, and spectral shape changes are treated as systematic uncertainty.

\subsection{Surface $^{210}$Pb}
\label{sec:surf}
\begin{figure}[t]
    \centering
    \centerline{\includegraphics[width=0.95\linewidth]{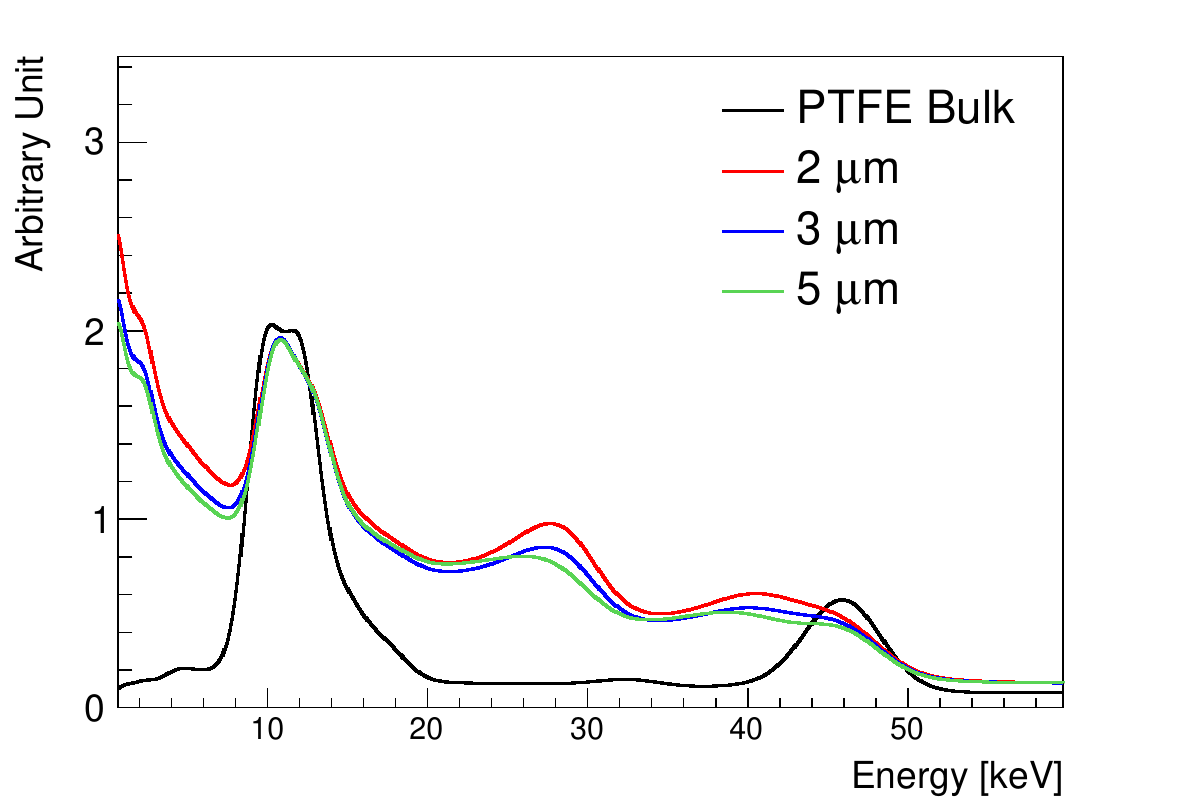}}
    \caption{Comparison of the MC spectrum of $^{210}$Pb on the PTFE surface with $^{210}$Pb in the PTFE bulk. The exponential mean depths of 2, 3, and 5\,$\mu$m for $^{210}$Pb on the PTFE surface are inferred from the alpha spectrum modeling~\cite{COSINE-100:alpha}.}
    \label{fig:tefsurf}  
\end{figure}
There is a low-energy background contribution due to the beta decay of $^{210}$Pb ($\beta^{-}$, 63.5\,keV, 22.3 years) on the crystal surface. This occurs because when the $^{210}$Bi-excited state undergoes de-excitation, it emits low-energy electrons and $\gamma$/X-rays. Some of these emissions can partially escape the crystal, depending on their depth relative to the surface.
This contribution was evaluated by measuring the surface depth profile, revealing its influence at various depths~\cite{Yu:2021depth}. 
The impact of surface depth-dependent backgrounds is considered in the current modeling process. 

Furthermore, a study of the alpha spectrum modeling~\cite{COSINE-100:alpha} found that the continuum below 2.2~MeV is caused by the presence of $^{210}$Po on the surface of the PTFE reflector surrounding the NaI(Tl) crystal. Based on the alpha modeling, it was determined that the PTFE surface contamination has exponential depth profiles with mean depths of 2--5\,$\mu$m.
The differences in their spectral shapes at various depths are illustrated in Fig.~\ref{fig:tefsurf}.
The peaks around 10\,keV and 46\,keV, which are prominent for the PTFE bulk contamination, result from X-rays and the 46.5\,keV $\gamma$-ray emitted from the decay of $^{210}$Pb, respectively. Conversion electrons contribute peaks around 28\,keV and 40\,keV, while beta electrons contribute to a continuum between peaks at low energy. Depending on their penetration depth, conversion electrons can escape partially from the PTFE surface and deposit energy in the crystal, similar to the crystal surface contaminant. These impacts on backgrounds are considered in the modeling process.

\subsection{K-shell X-ray emission from copper}
\begin{figure}[t]
    \centering
    \centerline{\includegraphics[width=0.95\linewidth]{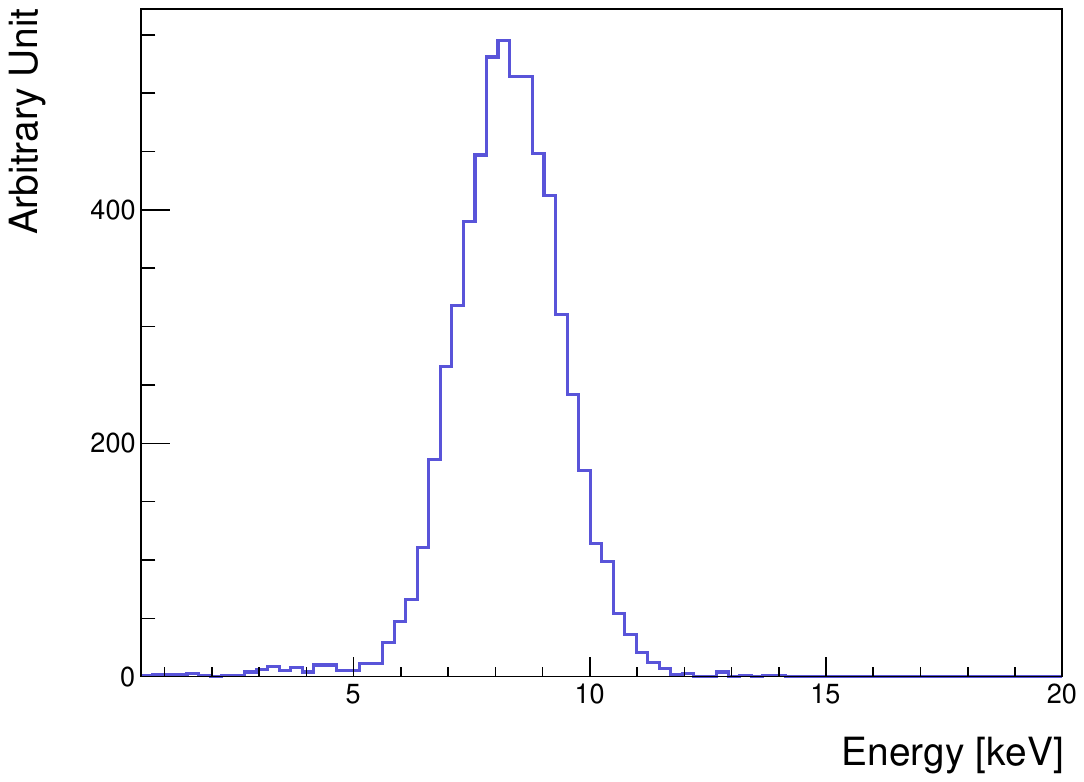}}
    \caption{The simulated energy spectrum of copper K-shell X-ray emission by shooting 12\,keV $\gamma$-ray, after applying detector smearing effect summarized in Fig.~\ref{fig:calibfunc}.}
    \label{fig:8keVSim}  
\end{figure}

To improve the accuracy of background modeling at low energies around 8\,keV, where the MC spectrum does not fully describe the data, we introduce the X-rays induced by the interaction of external backgrounds with materials such as copper. The NaI(Tl) crystals are encapsulated by a 1.5\,mm-thick copper cylinder, and in the presence of these backgrounds, X-ray emission is possible.
Specifically, there are X-ray emission lines at 8.038 and 8.905\,keV, corresponding to the mean energies of K$_{\alpha}$ and K$_{\beta}$ transitions of copper. 
The copper surface may be contaminated by a radioactive substance from outside, such as $^{222}$Rn, resulting in the implantation of $^{210}$Pb in its surface, similar to how the PTFE surface is contaminated with $^{210}$Pb. $\gamma$/X-rays emitted during the decay of the surface $^{210}$Pb can then induce X-ray emissions in copper.
We simulated this energy spectrum in a copper encapsulator by irradiating a 12\,keV $\gamma$-ray, similar to the X-ray energy from the decay of $^{210}$Pb, as shown in Fig.~\ref{fig:8keVSim}, and incorporated it into the modeling process.

\subsection{Background models}
\begin{figure*}[ht]
    \centering
    \centerline{\includegraphics[width=0.95\linewidth]{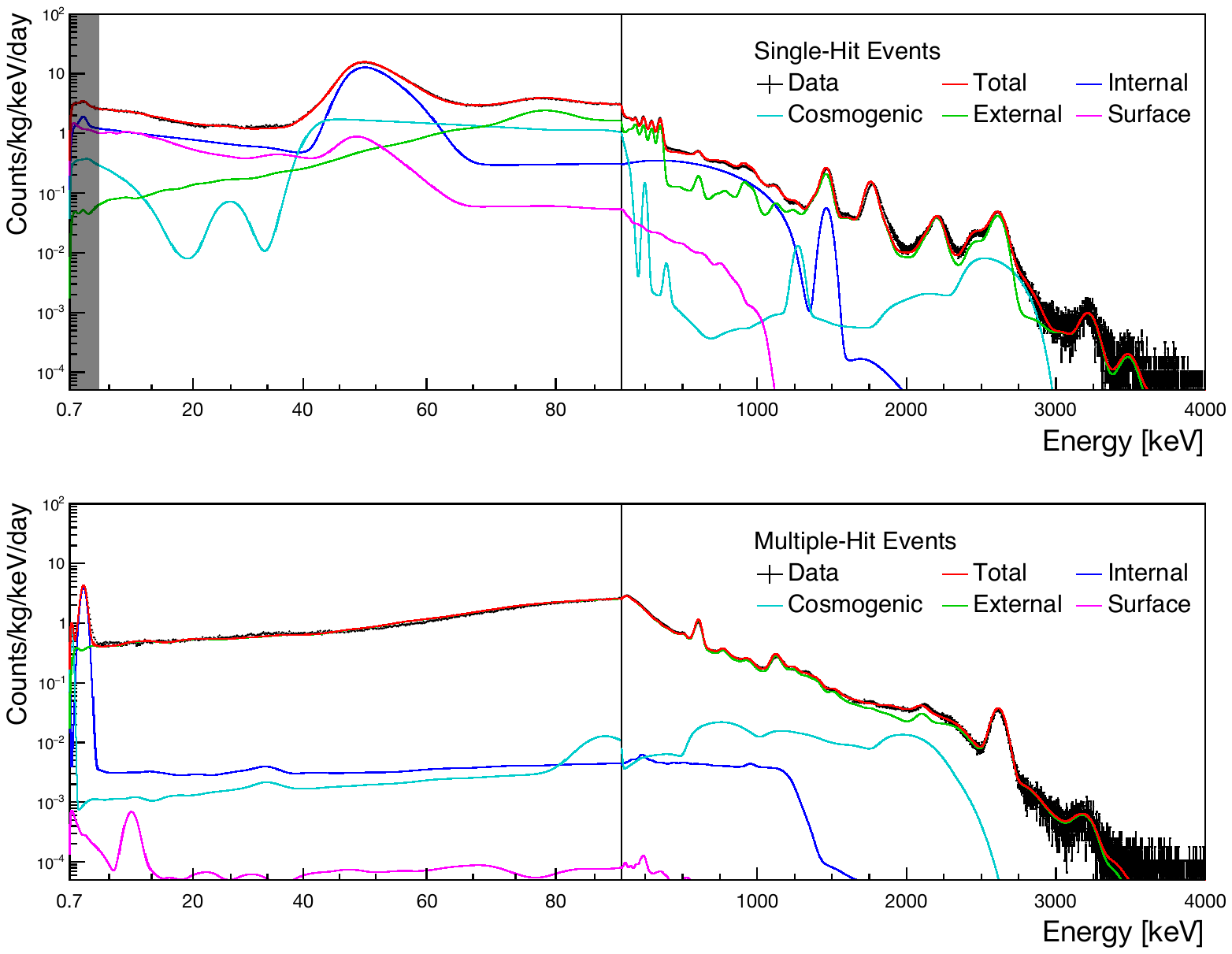}}
    \caption{Background modeling result for C2. The fitting region starts at 6~keV for single-hit events and 0.7 keV for multiple-hit events. The shaded region is excluded from the fitting, and the MC spectrum is extrapolated from the higher energy fitting results.  
    }
    \label{fig:ModelingResult}  
\end{figure*}

\begin{figure*}[ht]
    \centering
    \centerline{\includegraphics[width=0.65\linewidth]{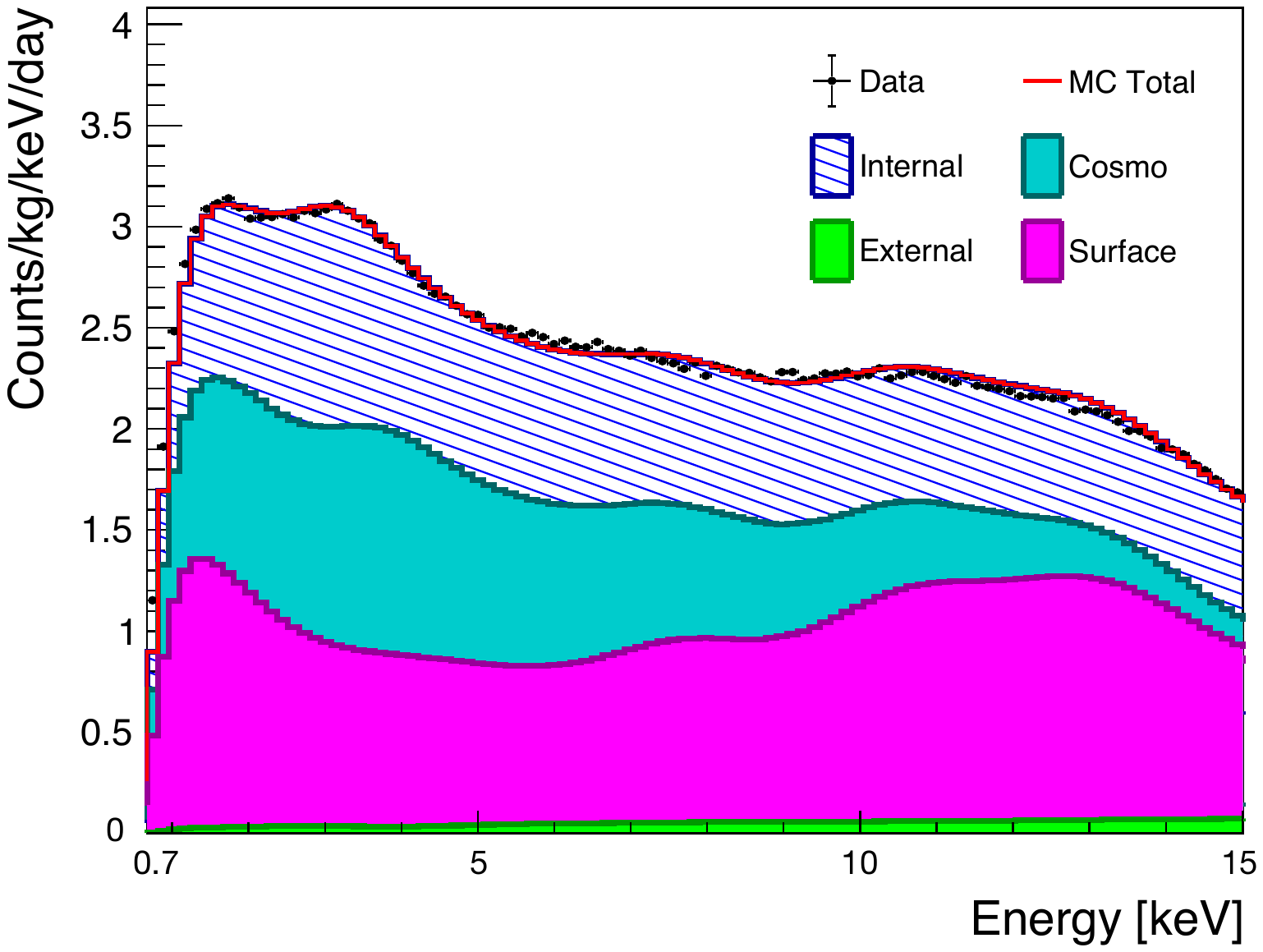}}
    \caption{Energy spectrum averaged over 5 crystals for single-hit events. Values below 6~keV are extrapolated from fitting results in other energy regions.}
    \label{fig:extrapolated}  
\end{figure*}

The data and Monte Carlo (MC) spectra are divided into four channels: single-hit low- and high-energy, multiple-hit low- and high-energy channels. Each channel utilizes an extended fitting energy range: the single-hit low-energy channel uses an energy range of [6, 90]\,keV, the multiple-hit low-energy channel utilizes a fitting region of [0.7, 90]\,keV, and both high-energy channels employ a fitting range of [90, 4000]\,keV.
The fitter simultaneously fits each channel to determine the fractional activity of the background sources. 

Figure~\ref{fig:ModelingResult} displays the background modeling results for crystal C2 as a representative. The top two plots show the single-hit low- and high-energy spectra, while the bottom two plots present the multiple-hit spectra. Consequently, the high-energy spectrum shows good agreement between data and MC across the entire energy region. Similarly, the low-energy spectrum is well-modeled, with no significant discrepancies observed. The WIMP search region of interest, which is grey-shaded and spans from 0.7 to 6.0\,keV, is excluded from the likelihood fitting procedure and represents an extrapolation based on the modeling results. The average of the extrapolated MC spectra for five crystals can be found in Fig.~\ref{fig:extrapolated}.

The single-hit low-energy spectrum shown in Fig.~\ref{fig:extrapolated} is mainly affected by surface $^{210}$Pb, cosmogenic isotope $^{3}$H, and internal radioactive isotopes $^{40}$K and $^{210}$Pb, which are represented by different colors (magenta, blue, and hatched area, respectively). 
The $^{210}$Pb contamination on the NaI(Tl) crystal surface, PTFE reflector, and the copper encapsulator's surface significantly influences the low-energy region.
According to the results in Table 1, surface contamination contributes to more than one-third of the total counts in most crystals. Additionally, there is a contribution from internal radioactive isotopes $^{210}$Pb and $^{40}$K, along with cosmogenic components such as $^{3}$H, $^{22}$Na, and $^{109}$Cd. The measurements on them are well-established and constrain their fractional activities in the modeling, showing good agreement with the data.

\begin{table*}[h]
\centering
\resizebox{\textwidth}{!}{
\begin{tabular}{@{}ccccccc@{}}
\toprule
\multicolumn{2}{c}{Unit [Counts/kg/keV/day]}     & Crystal 2     & Crystal 3   & Crystal 4         & Crystal 6         & Crystal 7         \\
\toprule
\multicolumn{2}{c}{Data}     
& 3.241 $\pm$ 0.007   &  3.363 $\pm$ 0.008  & 3.177 $\pm$ 0.005   &  2.830 $\pm$ 0.006   & 2.864 $\pm$ 0.006    \\
\midrule
\multicolumn{2}{c}{Total Simulation}   
&   3.335 $\pm$ 0.158 & 3.526 $\pm$ 0.041 &  3.129 $\pm$ 0.044 & 2.945 $\pm$ 0.047   &  2.937 $\pm$ 0.048   \\
\midrule
\multirow{3}{*}{Internal}    
& $^{210}\text{Pb}$
& 1.297 $\pm$ 0.005  & 0.375 $\pm$ 0.002 & 0.479 $\pm$ 0.002  & 1.115 $\pm$ 0.003 & 1.073 $\pm$ 0.003        \\
& $^{40}\text{K}$     
& 0.221 $\pm$ 0.002  & 0.098 $\pm$ 0.001 & 0.106 $\pm$ 0.001 & 0.044 $\pm$ 0.001  & 0.047 $\pm$ 0.001  \\
& {Others} 
& 0.0011 $\pm$ 0.0001 & 0.0004 $\pm$ 0.00004 & 0.0008 $\pm$ 0.0001 & 0.0005 $\pm$ 0.00003 &  0.0005 $\pm$ 0.00003 \\
\midrule
\multirow{2}{*}{Surface}  
& Crystal surface $^{210}$Pb   
& 0.683 $\pm$ 0.060  & 1.225 $\pm$ 0.012 &  0.000 $\pm$ 0.004 & 1.188 $\pm$ 0.015   & 1.242 $\pm$ 0.015       \\
& PTFE $^{210}$Pb              
& 0.034 $\pm$ 0.002  & 0.090 $\pm$ 0.002 &  0.031 $\pm$ 0.003  & 0.045 $\pm$ 0.003  & 0.026 $\pm$ 0.003     \\
\midrule
\multirow{5}{*}{Cosmogenics} 
& $^{3}\text{H}$  
&  0.341 $\pm$ 0.030  & 1.618 $\pm$ 0.018 &  1.629 $\pm$ 0.018  &  0.463 $\pm$ 0.020   &  0.451 $\pm$ 0.021     \\
 & $^{22}\text{Na}$                   
 & 0.0145 $\pm$ 0.0001  & 0.012 $\pm$ 0.0003 & 0.021 $\pm$ 0.0001 & 0.016 $\pm$ 0.0004  & 0.020 $\pm$ 0.0003       \\
& $^{109}\text{Cd}$                      
&  0.0061 $\pm$ 0.0035  & 0.043 $\pm$ 0.002 & 0.106 $\pm$ 0.002  & 0.011 $\pm$ 0.0002 &  0.011 $\pm$ 0.0005        \\
& $^{121m}\text{Te}$                     
& 0.0000 $\pm$ 0.0000   &  0.0021 $\pm$ 0.0001 & 0.0040 $\pm$ 0.0001 & 0.0025 $\pm$ 0.0001 &  0.0023 $\pm$ 0.0001       \\
 & Others           
 & 0.0098 $\pm$ 0.0008 &   0.0061 $\pm$ 0.0004 &  0.0178 $\pm$ 0.0004  & 0.0143 $\pm$ 0.0008  & 0.0115 $\pm$ 0.0004   \\
\midrule
\multicolumn{2}{c}{External}      
& 0.054 $\pm$ 0.005    &  0.058 $\pm$ 0.004  &  0.036 $\pm$ 0.003 &  0.045 $\pm$ 0.004 &  0.052 $\pm$ 0.004        \\
\bottomrule
\end{tabular}
}
\caption{Single-hit event rate of major background sources and data in the 0.7 to 6\,keV range.}
\label{table:bkgresult}
\end{table*}

\section{Conclusion} 
\label{sec:conc}
The COSINE-100 project has gathered data from October 21, 2016, to March 14, 2023. This study presents the outcomes of background modeling using a dataset spanning 2.82 years with a 0.7 keV energy threshold. The new analysis incorporates a more precise treatment of the non-proportional energy response of the NaI(Tl) detectors, taking into account both the low- and high-energy channels. This improved understanding allows for a more accurate assessment of the $^{22}$Na 0.87\,keV peak in the low-energy channel, as well as the multi-gamma energy deposits from $^{22}$Na and $^{208}$Tl around 2.8~MeV and above 3~MeV in the high-energy channel, setting the upper energy threshold of the modeling at 4~MeV. Additionally, the depth profile of surface $^{210}$Pb contamination in the PTFE reflector is included in the enhanced background modeling based on the study of the alpha background modeling. With the improved understanding of the surface $^{210}$Pb in the PTFE reflector, it is found that it can affect the low-energy spectrum as much as the crystal surface $^{210}$Pb contamination. 
The improved background modeling explained the overall energy spectrum and accurately represented the extrapolated WIMP ROI (single-hit energy of 0.7--6~keV) using the MC spectrum for the background. In the extrapolated WIMP search ROI, the background rate was measured as 3.167~$\pm$~0.022~Counts/kg/keV/day in 5 crystals on average, with the dominating background modeled as internal $^{3}$H, $^{210}$Pb, and surface-contaminating $^{210}$Pb.

   
\section*{Acknowledgments}
\sloppy 
We thank the Korea Hydro and Nuclear Power (KHNP) Company for providing underground laboratory space at Yangyang and the IBS Research Solution Center (RSC) for providing high performance computing resources. 
This work is supported by:  the Institute for Basic Science (IBS) under project code IBS-R016-A1,  NRF-2021R1A2C3010989, NRF-2021R1A2C1013761 and RS-2024-00356960,
NFEC (No. 2019R1A6C1010027) and NRF (No. 2021R1I1A3041453),
Republic of Korea;
NSF Grants No. PHY-1913742, United States; 
STFC Grant ST/N000277/1 and ST/K001337/1, United Kingdom;
Grant No. 2021/06743-1, 2022/12002-7 and 2022/13293-5 FAPESP, CAPES Finance Code 001, CNPq 304658/2023-5, Brazil;
UM grant No. 4.4.594/UN32.14.1/LT/2024, Indonesia.

\bibliographystyle{elsarticle-num} 
\bibliography{biblio.bib}

\begin{thebibliography}{10}
\expandafter\ifx\csname url\endcsname\relax
  \def\url#1{\texttt{#1}}\fi
\expandafter\ifx\csname urlprefix\endcsname\relax\def\urlprefix{URL }\fi
\expandafter\ifx\csname href\endcsname\relax
  \def\href#1#2{#2} \def\path#1{#1}\fi

\bibitem{Gaitskell:2004gd}
R.~J. Gaitskell, {Direct detection of dark matter}, Ann. Rev. Nucl. Part. Sci.
  54 (2004) 315--359.
\newblock \href {https://doi.org/10.1146/annurev.nucl.54.070103.181244}
  {\path{doi:10.1146/annurev.nucl.54.070103.181244}}.

\bibitem{particle2020review}
P.~D. Group, P.~Zyla, R.~Barnett, J.~Beringer, O.~Dahl, D.~Dwyer, D.~Groom,
  C.-J. Lin, K.~Lugovsky, E.~Pianori, et~al., {Review of particle physics},
  Progress of theoretical and experimental physics 2020~(8) (2020) 083C01.

\bibitem{DAMA:2008jlt}
R.~Bernabei, et~al., {First results from DAMA/LIBRA and the combined results
  with DAMA/NaI}, Eur. Phys. J. C 56 (2008) 333--355.
\newblock \href {http://arxiv.org/abs/0804.2741} {\path{arXiv:0804.2741}},
  \href {https://doi.org/10.1140/epjc/s10052-008-0662-y}
  {\path{doi:10.1140/epjc/s10052-008-0662-y}}.

\bibitem{DAMA:2010gpn}
R.~Bernabei, et~al., {New results from DAMA/LIBRA}, Eur. Phys. J. C 67 (2010)
  39--49.
\newblock \href {http://arxiv.org/abs/1002.1028} {\path{arXiv:1002.1028}},
  \href {https://doi.org/10.1140/epjc/s10052-010-1303-9}
  {\path{doi:10.1140/epjc/s10052-010-1303-9}}.

\bibitem{Bernabei:2013xsa}
R.~Bernabei, et~al., {Final model independent result of DAMA/LIBRA-phase1},
  Eur. Phys. J. C 73 (2013) 2648.
\newblock \href {http://arxiv.org/abs/1308.5109} {\path{arXiv:1308.5109}},
  \href {https://doi.org/10.1140/epjc/s10052-013-2648-7}
  {\path{doi:10.1140/epjc/s10052-013-2648-7}}.

\bibitem{Bernabei:2018jrt}
R.~Bernabei, et~al., {First model independent results from DAMA/LIBRA-phase2},
  Nucl. Phys. Atom. Energy 19~(4) (2018) 307--325.
\newblock \href {http://arxiv.org/abs/1805.10486} {\path{arXiv:1805.10486}},
  \href {https://doi.org/10.15407/jnpae2018.04.307}
  {\path{doi:10.15407/jnpae2018.04.307}}.

\bibitem{bernabei2023dark}
R.~Bernabei, P.~Belli, F.~Cappella, V.~Caracciolo, R.~Cerulli, C.~Dai,
  A.~d’Angelo, A.~Incicchitti, A.~Leoncini, X.~Ma, et~al., {Dark Matter with
  DAMA/LIBRA and its perspectives}, in: Journal of Physics: Conference Series,
  Vol. 2586, IOP Publishing, 2023, p. 012096.

\bibitem{adhikari2021strong}
G.~Adhikari, E.~B. de~Souza, N.~Carlin, J.~J. Choi, S.~Choi, M.~Djamal, A.~C.
  Ezeribe, L.~E. Fran{\c{c}}a, C.~H. Ha, I.~S. Hahn, et~al., {Strong
  constraints from COSINE-100 on the DAMA dark matter results using the same
  sodium iodide target}, Science advances 7~(46) (2021) eabk2699.

\bibitem{adhikari2022three}
G.~Adhikari, E.~Barbosa~de Souza, N.~Carlin, J.~Choi, S.~Choi, A.~Ezeribe,
  L.~Fran{\c{c}}a, C.~Ha, I.~Hahn, S.~Hollick, et~al., {Three-year annual
  modulation search with COSINE-100}, Physical Review D 106~(5) (2022) 052005.

\bibitem{adhikari2023search}
G.~Adhikari, N.~Carlin, J.~J. Choi, S.~Choi, A.~Ezeribe, L.~Fran{\c{c}}a,
  C.~Ha, I.~Hahn, S.~Hollick, E.~Jeon, et~al., {Search for boosted dark matter
  in COSINE-100}, Physical review letters 131~(20) (2023) 201802.

\bibitem{COSINE-100:nPR}
S.~Lee, et~al., {Nonproportionality of NaI (Tl) Scintillation Detector for Dark
  Matter Search Experiments}, arXiv preprint arXiv:2401.07462 (2024).

\bibitem{COSINE-100:alpha}
G.~Adhikari, N.~Carlin, D.~Cavalcante, J.~Cho, J.~Choi, S.~Choi, A.~Ezeribe,
  L.~Fran{\c{c}}a, C.~Ha, I.~Hahn, et~al., {Alpha backgrounds in NaI (Tl)
  crystals of COSINE-100}, Astroparticle Physics (2024) 102945.

\bibitem{Adhikari:2017esn}
G.~Adhikari, et~al., {Initial Performance of the COSINE-100 Experiment}, Eur.
  Phys. J. C 78~(2) (2018) 107.
\newblock \href {http://arxiv.org/abs/1710.05299} {\path{arXiv:1710.05299}},
  \href {https://doi.org/10.1140/epjc/s10052-018-5590-x}
  {\path{doi:10.1140/epjc/s10052-018-5590-x}}.

\bibitem{COSINE-100:2021mrq}
G.~Adhikari, et~al., {Background modeling for dark matter search with 1.7~years
  of COSINE-100 data}, Eur. Phys. J. C 81~(9) (2021) 837.
\newblock \href {http://arxiv.org/abs/2101.11377} {\path{arXiv:2101.11377}},
  \href {https://doi.org/10.1140/epjc/s10052-021-09564-0}
  {\path{doi:10.1140/epjc/s10052-021-09564-0}}.

\bibitem{COSINE-100:2020wrv}
G.~Adhikari, et~al., {Lowering the energy threshold in COSINE-100 dark matter
  searches}, Astropart. Phys. 130 (2021) 102581.
\newblock \href {http://arxiv.org/abs/2005.13784} {\path{arXiv:2005.13784}},
  \href {https://doi.org/10.1016/j.astropartphys.2021.102581}
  {\path{doi:10.1016/j.astropartphys.2021.102581}}.

\bibitem{COSINE-100:eventselection-2024}
G.~Yu, et~al., {Lowering threshold of NaI(Tl) scintillator to 0.7 keVee in the
  COSINE-100 experiment}, in the preparation.

\bibitem{engelkemeir1956nonlinear}
D.~Engelkemeir, {Nonlinear response of NaI (Tl) to photons}, Review of
  Scientific Instruments 27~(8) (1956) 589--591.

\bibitem{jones1962nonproportional}
T.~H. Jones, {The nonproportional response of a NaI (Tl) crystal to diffracted
  X rays}, Nuclear Instruments and Methods 15~(1) (1962) 55--58.

\bibitem{leutz1997scintillation}
H.~Leutz, C.~D'Ambrosio, {On the scintillation response of NaI (TI)-crystals},
  IEEE Transactions on Nuclear Science 44~(2) (1997) 190--193.

\bibitem{collinson1963fluorescent}
A.~Collinson, R.~Hill, {The fluorescent response of NaI (Tl) and CsI (Tl) to X
  rays and $\gamma$ rays}, Proceedings of the Physical Society 81~(5) (1963)
  883.

\bibitem{khodyuk2010nonproportional}
I.~Khodyuk, P.~Rodnyi, P.~Dorenbos, {Nonproportional scintillation response of
  NaI: Tl to low energy x-ray photons and electrons}, Journal of Applied
  Physics 107~(11) (2010).

\bibitem{moses2012origins}
W.~Moses, G.~Bizarri, R.~T. Williams, S.~Payne, A.~Vasil'Ev, J.~Singh, Q.~Li,
  J.~Grim, W.-S. Choong, {The origins of scintillator non-proportionality},
  IEEE Transactions on Nuclear Science 59~(5) (2012) 2038--2044.

\bibitem{choi2024waveform}
J.~Choi, C.~Ha, E.~Jeon, K.~Kim, S.~Kim, Y.~Kim, Y.~Ko, B.~Koh, H.~Lee, S.~Lee,
  et~al., {Waveform simulation for scintillation characteristics of NaI (Tl)
  crystal}, Nuclear Instruments and Methods in Physics Research Section A:
  Accelerators, Spectrometers, Detectors and Associated Equipment 1065 (2024)
  169489.

\bibitem{DEVARE1963253}
H.~Devare, P.~Tandon, Effect of the non-linear response of nai(tl) on the
  single crystal summing spectra, Nuclear Instruments and Methods 22 (1963)
  253--255.

\bibitem{GEANT4:2002zbu}
S.~Agostinelli, et~al., {GEANT4--a simulation toolkit}, Nucl. Instrum. Meth. A
  506 (2003) 250--303.
\newblock \href {https://doi.org/10.1016/S0168-9002(03)01368-8}
  {\path{doi:10.1016/S0168-9002(03)01368-8}}.

\bibitem{Allison:2006ve}
J.~Allison, et~al., {Geant4 developments and applications}, IEEE Trans. Nucl.
  Sci. 53 (2006) 270.
\newblock \href {https://doi.org/10.1109/TNS.2006.869826}
  {\path{doi:10.1109/TNS.2006.869826}}.

\bibitem{Allison:2016lfl}
J.~Allison, et~al., {Recent developments in Geant4}, Nucl. Instrum. Meth. A 835
  (2016) 186--225.
\newblock \href {https://doi.org/10.1016/j.nima.2016.06.125}
  {\path{doi:10.1016/j.nima.2016.06.125}}.

\bibitem{COSINE-100:2018tfl}
P.~Adhikari, et~al., {Background model for the NaI(Tl) crystals in COSINE-100},
  Eur. Phys. J. C 78 (2018) 490.
\newblock \href {http://arxiv.org/abs/1804.05167} {\path{arXiv:1804.05167}},
  \href {https://doi.org/10.1140/epjc/s10052-018-5970-2}
  {\path{doi:10.1140/epjc/s10052-018-5970-2}}.

\bibitem{Yu:2021depth}
G.~Yu, C.~Ha, E.~J. Jeon, K.~Kim, N.~Y. Kim, Y.~Kim, H.~S. Lee, H.~Park,
  C.~Rott, {Depth profile study of 210Pb in the surface of an NaI (Tl)
  crystal}, Astroparticle Physics 126 (2021) 102518.

\end{thebibliography}

\end{document}